# InteractiveEdu: An Open-source Interactive Floor for Exergame as a Learning Platform

Everson Borges da Rosa[1][a], Michel Albonico[1][b], and Paulo Júnior Varela[1][c]

[1]*Federal University of Technology, Francisco Beltrão, Brazil*
michelalbonico@utfpr.edu.br

Keywords: Active methodologies; Learning Platform; Exergame; Interactive Floor.

Abstract: Children tend to be constantly exposed to technologies, such as smartphones, tablets, and gaming consoles, drawn by the interactive and visually stimulating nature of digital platforms. Thus, integrating the teaching process with technological gadgets may enhance engagement and foster interactive learning experiences, besides equipping students with the digital skills for today's increasingly technology-driven world. The main goal of this work is to provide an open-source and manageable tool that teachers can use as an everyday activity and as an exergame. For this, we present a prototype of an interactive platform that students use to answer a quiz by moving to segments available on an interactive floor. All the platform design and implementation directions are publicly available.

## 1 INTRODUCTION

Currently, technological resources are present in various everyday situations, disseminated through increasingly comprehensive means of communication, and through the generalization of urban lifestyles, different types of social relationships emerge, leading to a new understanding of the world. Technology is integrated into our lives to such an extent that even in the simplest activities we may no longer perceive them as anything other than natural occurrences [Kenski, 2003]. After twenty years, such an affirmation became a reality with technology changing the way we communicate in the modern world, even influencing our consumption behavior, where the local loses space for e-commerce.

In education, technology has also become a reality, despite "schools proving slower to change their lesson plans than they were to fit computers in the classroom" [Livingstone, 2015, Watson et al., 2013]. The use of technology as a teaching tool can benefit teachers who as a mediator of knowledge, face a daily challenge to spark students' interest in the content of their lessons amidst a plethora of resources capable of diverting students' attention [Bulman and Fairlie, 2016]. Different initiatives have been seeking to reconcile technology with learning, including educational robotics [Tzagkaraki et al., 2021], which has been growing in recent years through the well-known LEGO [Souza et al., 2018] and Arduino [Slomski et al., 2022, Perenc et al., 2019] platforms.

Another key aspect to be considered is technology as a booster of sedentary behavior [Dow and Reed, 2015], where its extensive use could be harmful. Children and young people are spending more and more hours sitting in front of computers, smartphones, and other devices. Even when used for educational purposes, there is often little interaction with the physical environment outside the virtual realm. In this context, exergames (games that require bodily movements to be played) could be useful tools for schools to promote physical activity and combat obesity in adolescents [Garde et al., 2016, Staiano et al., 2012].

Given the current reality, it is necessary to adopt reactive measures, which help students and teachers reconcile the use of technology as a facilitator while keeping children physically active. In **this work**, we propose a learning platform with an interactive floor that communicates with a Web-based quiz. The platform also provides a mobile app for teachers to register a database of questions with optional answers. With our interactive learning platform, it is possible to carry out educational assignments combined with physical activities (as an exergame), in which students, moving over the interactive floor, respond to the questionnaires presented on the web interface.

[a] https://orcid.org/0000-0000-0000-0000

[b] https://orcid.org/0000-0000-0000-0000

[c] https://orcid.org/0000-0000-0000-0000

## 2 BACKGROUND

In this section, we briefly discuss the technologies and methodologies necessary to comprehend the proposed interactive learning platform.

### 2.1 Active Methodologies

We learn in countless ways from birth when facing challenges in all fields (personal, social, professional), which empowers us to alter the reality in which we live, not just adapt to it [Freire, 1996, Cintra and Fedel, 2019]. As human beings, we learn through concrete situations that can maximize the inductive process, while we also acquire knowledge from ideas or theories that can subsequently be tested concretely, thereby disseminating the deductive process [Bacich and Moran, 2018].

In active learning methodology, the learner plays a central role and is primarily responsible for the learning method [Felder and Brent, 2009]. Such a teaching approach aims at stimulating the development of the ability to assimilate content independently and participatively. Active methodologies play a significant role in education, enabling the possibility of classes with more meaningful learning for digital culture students who have a different understanding of teaching and learning compared to previous generations [Bacich and Moran, 2018].

#### 2.1.1 Exergames

Exercise games (i.e., *exergames*) are a branch of active methodology that joins technology and physical activity, where the player stays physically active during the game [Song et al., 2011]. For this, it is common to use motion-sensing technologies, such as accelerometers, cameras, or wearable sensors, to monitor and respond to the player's physical movements in real-time.

### 2.2 Educational Robotics

Educational Robotics, also known as Pedagogical Robotics, has emerged as a comprehensive resource in the field of education, being also aligned with the development of skills highly demanded by future professionals [Daniela and Lytras, 2019]. It refers to the learning environment in which the teacher teaches the student assembly, automation, and control of mechanical devices that can be controlled by the computer is called Pedagogical Robotics or Educational Robotics [Silva, 2009].

Educational Robotics involves a process of motivation, collaboration, construction, and reconstruction. The robot, as a technological element, encompasses a series of scientific concepts whose basic principles are addressed by the school. Furthermore, robots stimulate children's imagination, creating new forms of interaction [Silva, 2009].

### 2.3 Arduino

Alongside the demand for technologies that have been spreading recently, Arduino emerged as an electronic low-cost and generic prototyping platform for implementing interactive principles and developing system control [López-Rodríguez and Cuesta, 2016]. Figure 1 illustrates the Arduino UNO, a popular board provided with starting robotic kits sold on the Internet. It mainly features a microcontroller, and a wide array of input and output pins for interfacing with sensors, actuators, and other electronic components.

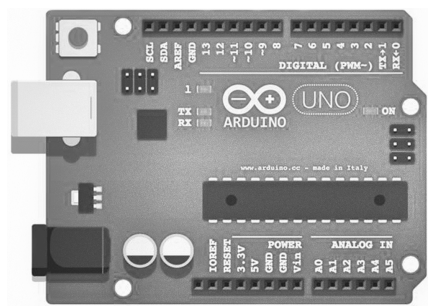

Figure 1: Arduino UNO board.

Arduino offers the possibility of easily integrating different electronic circuits, from sensors to actuators [Aguiar, 2018]. With its open-source nature, it incurs no licensing or copyright costs, making it a democratic tool that can benefit educational applications. Arduino maintainers also keep a hub with diverse projects that are publicly available, which can be replicated or improved by the community [Melo, 2012].

## 3 PLATFORM DESIGN

In this section, we present the design of the proposed interactive learning platform.

### 3.1 System Semantic

Figure 2 presents the ontology behind the Interactive Floor prototype, depicting the main components of the system and their interactions. *Interactive Floor* is the central component as a product of all the other component interactions. The *app* corresponds to a smartphone application used by the teacher to manage his/her profile together with *questions* and *answers*. Note that *questions* and *answers* are directly related, which means that for each *question* we must register corresponding *answers*. All the registered data is stored in a *database*, which serves the *web (quiz) server*. The *web (quiz) server* also interfaces with *Arduino* and *computer* devices, feeding the *computer* with the current quiz question and related answers, and receiving the stimulus from the students with the interactive *floor* (connected to the *Arduino* component). The *computer* presents the *Quiz* question and answer options to students on its screen (which can also be projected). All the communication happens over well-known Web-based protocols, which makes the proposed system modular and portable to different educational settings.

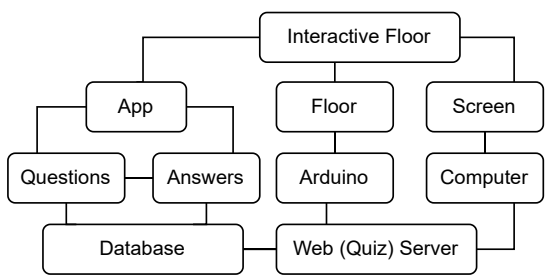

Figure 2: Interactive Edu ontology.

## 3.2 System Requirements

From the system ontology, we can extract the main functional requirements that may be addressed in the project:
*Register teacher*: the system must provide an area for teachers to register themselves as users.
*Authenticate teacher*: the system must provide a login screen where the user will enter their credentials before being granted access to further features of the system.
*Manage questions*: the system must provide an area for teachers to manage the questions (add, edit, and remove) the students are later exposed to.
*Manage answers*: the system must allow teachers to register answers for the registered questions, classifying them as correct or wrong.
*Show Quiz*: the system must provide a quiz interface for the students to interact with registered questions. This must be displayed as a screen.
*Interact with Quiz*: the system must provide a way to connect the quiz with an interactive interface.

Figure 3 illustrates the system requirements dependencies modeled as a Unified Modeling Language (UML) use case diagram. For the teacher to authenticate, he/she must have previously been registered to the system, and once authenticated, the teacher can manage both, questions and answers. The registered questions and related answers are then presented to a student on a screen in the form of a Quiz. Finally, the student interacts with the Quiz by moving to the correct position on the Interactive Floor, which interprets the stimulus as something similar to a mouse click, and sends the chosen answer to the Quiz server.

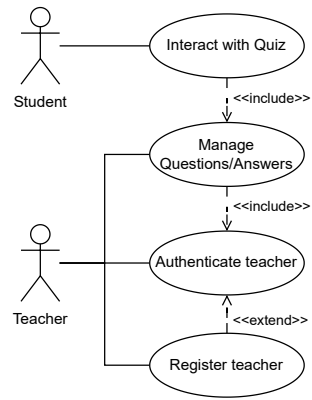

Figure 3: Requirements use case diagram.

## 3.3 Teacher Management System

To facilitate the management and control of the question database used by the interactive floor, we designed a mobile application (app) for smartphones. For granting access only to registered teachers, the app must provide an authentication interface. Then, after logging in, the teacher must be redirected to the options menu screen, where he/she can choose between adding new questions or viewing existing ones. An additional screen for the view of existing questions enables the teacher to remove questions.

We also design a database for storing the questions with the corresponding answer alternatives. Figure 4 illustrates the entity relationship diagram for such a database. In the diagram, we observe the multiplicity of answers for each question, where each answer is set as either correct or not (boolean data type).

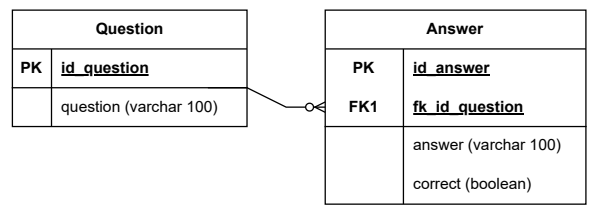

Figure 4: Entity Relationship Diagram of the questions and answers.

## 3.4 Quiz Server

The *Quiz server* is used to share the data stored in the questions database. Note that in the current version of the Interactive Edu, the server only provides data, without receiving any feedback. However, we designed it with a bidirectional communication, where the student answers can be computed and ranked in the future, which would be an extra stimulus for them to play with the tool. The *Quiz server* must also serve as a bridge between the different student interfaces (presented in the sequence), enabling them to keep exchanging data, such as interactions of the students with the interactive floor being sent to the screen where the questions are presented.

## 3.5 Student Interface

The students interact with two different interfaces: i) the Quiz screen, where they see sequential questions with optional answers; and ii) the Interactive Floor, where they must stimulate/jump to answer the questions. Both are presented in the sequence.

### 3.5.1 Quiz Screen

Figure 5 illustrates the design of the Quiz screen, which is chosen to be simple (only with necessary information) and colorful (to keep children's attention).

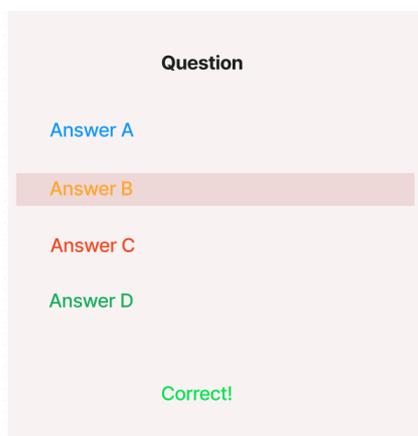

Figure 5: Sketch of the screen that presents the questions to the students.

The screen interface is designed to keep children's attention to the question and its answers. At the center, the question is posed in a bold and legible font, which serves as the focal point of the quiz. Directly beneath the question, four colorful answer options are displayed. Each answer option is presented as a clickable element, resembling playful buttons. They are presented in vibrant colors, making them visually appealing and distinct from one another. These colorful elements are linked to the colors in the proposed floor device (presented in the sequence).

Upon selecting an answer, a feedback message must be promptly displayed below the answer options. The feedback message must dynamically appear in a large, easy-to-read font at the center of the screen. If the chosen answer is correct, the feedback message must ppear in a cheerful green color, stating "Correct!". However, if the answer is incorrect, the message must appear in a contrasting red color, gently informing the child, "I'm sorry, but it is wrong!". This immediate feedback mechanism enhances the learning experience, providing instant reinforcement and guidance to the child.

### 3.5.2 Interactive Floor

We aim for the interactive floor to be modular and adaptative (others should be able to evolve it according to their needs), and be independent of the learning platform (should be connected to other platforms without much effort). Figure 6 illustrates the electronic circuit[1] of the Arduino connected to an electronic circuit that represents the interactive floor. In the figure, the areas of the mat are represented by push buttons. These buttons have the functionality to close or open the electrical circuit, which is responsible for indicating the position of the student on the mat. Each button is connected to an Arduino port, which corresponds to an answer displayed on the screen to the student.

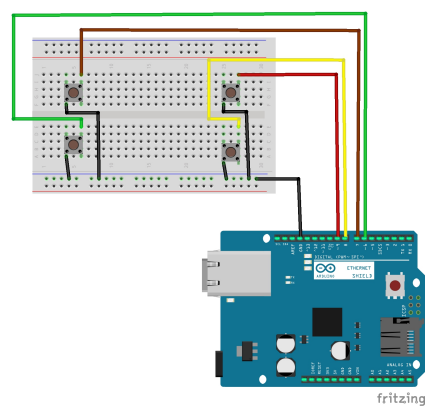

Figure 6: Sketch of the proposed Arduino's electronic circuit.

---

[1]Generated with Fritzing: https://fritzing.org/

## 4 PLATFORM IMPLEMENTATION

In this section, we give more details about the implementation of the proposed interactive learning platform and also provide some screens and pictures of the current version of the prototype.

### 4.1 Quiz Server

The *Quiz Server* is implemented with a MySQL DBMS, where the questions and other data are stored. In the current version, it only receives the new data from the teacher's management app, synchronizes both databases, and serves the student interfaces. For this, there are two basic services enabled in the *Quiz server*: i) a PHP Web page, where the application sends new data via POST request; ii) a WebSocket where student interfaces connect and retrieve the data from the database. For data exchange, the *Quiz server* relies on a JSON file.

### 4.2 Teacher's Management App

The teacher's management app is implemented with MIT App Inventor. Despite the restrictions in the user interface (UI), given the limitations of the app, the project is publicly available, which enables others to reuse it and update it according to their needs. Figure 7 presents the main screen of the app, with all the functions according to its design, with a few changes in the elements disposal. All the data is stored locally, in an *tinyDB*[2] database. In the screen where the questions are listed, there is a button to synchronize with the *Quiz server* database via the Hypertext Transfer Protocol (HTTP) POST method (given the restrictions of the MIT App Inventor platform).

### 4.3 Quiz Screen

The *Quiz screen* was developed as a simple Web page, which is directly connected to the *Quiz server* via a WebSocket connection. All the code is self-contained in a single HTML/JavaScript file. Since the Web page runs static code, there is no need for further software layers to be installed, making it portable to different machines, only requiring a recent Web browser to be installed.

Figure 8 illustrates a question as it is shown to the students. The result screen is clean as designed with colorful answers, and keeps the interaction area highlighted.

---
[2] https://tinydb.readthedocs.io/en/latest/

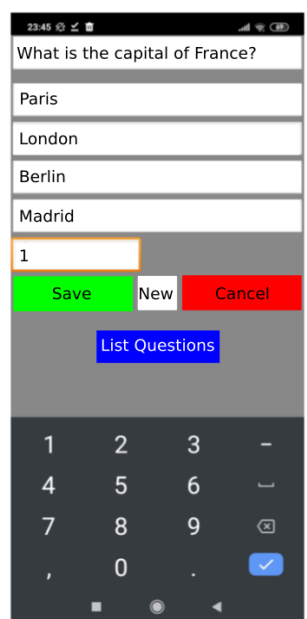

Figure 7: App main screen of the teacher's management app.

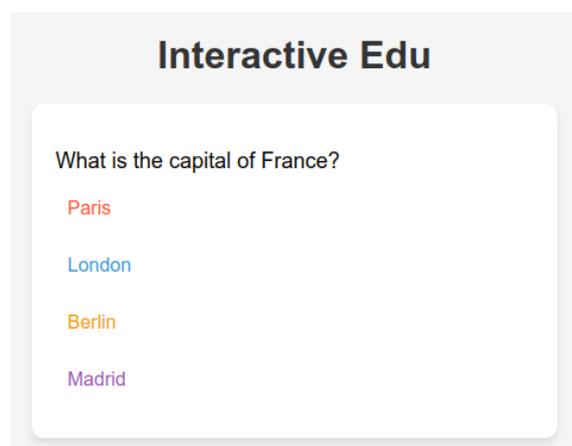

Figure 8: Quiz screen presenting the question and the optional answers.

Figure 10 illustrates the *Quiz screen* when a student selects a correct answer. We can observe that the command sent from the Arduino board is received as a stimulus similar to a mouse click and that in the case of the correct answer, a warning message in green is presented at the bottom of the Web page.

### 4.4 Interactive Floor

For the integration of the *interactive floor* with the *Quiz screen*, we rely on Arduino. Thus, the properly configured microcontroller became responsible for capturing the stimulus from the electrical circuit, processing the data obtained according to the estab-lished algorithm, and sending it to the Web interface where the questions are presented. For connecting the Arduino board to the internet network, we rely on an Ethernet Shield[3]. All the communication happens over a WebSocket connection[4] open with the *Quiz server*, which bridges the data exchange between the *Quiz screen* and the *interactive floor*, both clients of that WebSocket server. The implementation enables the interactive floor to be easily integrated into other platforms. Others can keep the *Quiz server* as a bridge, only requiring an extra service for transforming the received data into a format that meets the requirements of the further platform Application Programming Interface (API). This would require no adaptations to the interactive floor device and Arduino microcontroller.

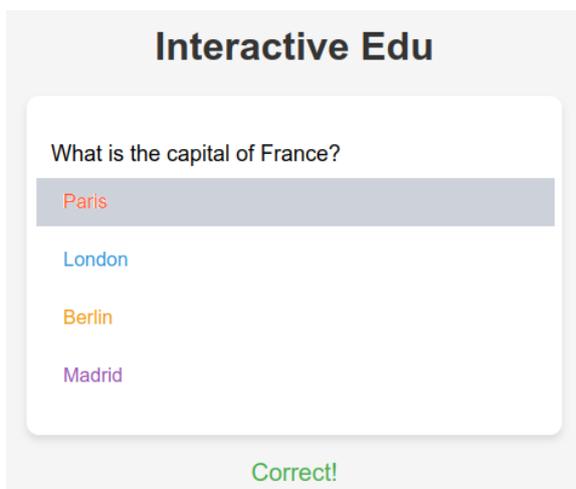

Figure 9: Quiz screen with the correct answer selected.

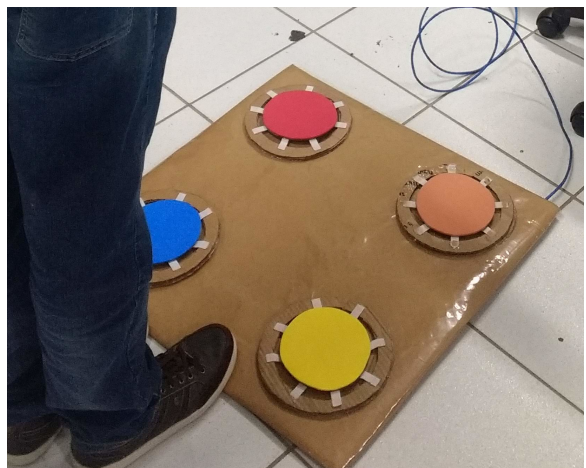

Figure 10: Interactive floor prototype.

## 5 DISCUSSION

The interactive floor for exergame as a learning platform described in the text can have several practical applications in educational settings. As a platform, we reinforce that it may be applied to:

*Engaging classroom activities*: teachers can use the proposed platform to create engaging classroom activities that combine physical movement with learning. This can help in keeping students active and focused during lessons.

*Promoting physical activity*: its exergame nature can be used to promote physical activity by incorporating movement into learning activities, teachers can encourage students to be more physically active throughout the school day.

*Developing digital skills*: given the openness and simplicity of the project, the students (depending on their age), can help teachers to build the interactive floor device. This can empower students to become familiar with basic electronics and sharpen their creativity.

The use of technological gadgets such as interactive floors for exergames can enhance student engagement and provide interactive learning experiences. However, it is crucial to critically consider the pedagogical implications of such approaches:

*Balance with traditional methods*: While technology can enhance learning experiences, it should not overshadow the importance of human interaction, critical thinking, and other essential skills that traditional education fosters. Teachers need to carefully design activities that leverage technology while also promoting a holistic educational experience.

*Keep promoting physical activity*: The potential for technology to promote sedentary behavior is a significant concern. It is well-known that children and young people are increasingly spending time in front of screens, which can have detrimental effects on their physical health. While exergames offer a solution by incorporating physical activity into learning, it is essential to ensure that such activities are not used as a substitute for regular exercise or outdoor play. It should only serve as a spark in terms of motivation, where teachers should be mindful of striking a balance between screen time and physical activity to promote overall well-being.

*Consider possible disparities in technology*: The accessibility to technology in education must be considered, especially in areas with social disadvantages. Not all students may have equal access to technology, which can create disparities in learning opportunities.

---

[3] https://docs.arduino.cc/retired/shields/arduino-ethernet-shield-without-poe-module/

[4] By using the WebSocket library: https://www.arduino.cc/reference/en/libraries/websockets/

Therefore, when a methodology such as the suggested is applied in a school scenario, it is important to ensure the activities are inclusive, and basic concepts are previously worked with the ones that are excluded from technological reality.

## 6 RELATED WORK

Since active methodologies have been recurrently used in the last years, we can find multiple studies that use interactive technologies for education. Here, we briefly discuss them and highlight the main differences from what is proposed in this paper.

Numerous works in the literature have proposed the use of technology in education. Singhal et al., 2012 uses an Augmented Reality (AR) for teaching Chemistry in an immersion environment. Mayer et al., 2022 proposes a gamification approach where the player goes through a Role-Playing Game (RPG) world and is asked to combine chemical elements. Sharaf. et al., 2023 use robots to introduce computational thinking to small kids (between 7-11 years old). da Silva and Schorr, 2023 propose a mobile app for supporting the literacy process of incoming students. Despite being diverse in the disciplines they address and the technologies they use, these works do not include bodily movements, and therefore, do not tackle the sedentarism challenge.

Other authors apply their research in the field of exergames. Gao et al., 2014 propose the use of casual exergames in school to engage adolescents to be more active physically. Their approach does not propose any new game. They use existing games associated with the Microsoft Kinect device. Macvean and Robertson, 2012 create a game they call Escape the Ghost inspired by the traditional children's game chase and catch. It synchronizes a screen game with movements in the real world. However, the game does not link to any school discipline. Compared to those works, our approach aggregates by being adaptative to different disciplines, and providing a full package with open-source code that can be replicated with low cost and little effort.

## 7 CONCLUSION

The development of this project was committed to fostering the relationship and communication between teachers and students during the learning process. The aim was to create a platform in which students would engage in physical activity alongside the use of technology while intending to generate greater interest in the content presented by the educator.

In the project, we prioritize open-source technologies and also make all the projects publicly available. Furthermore, the device prototype is implemented by using affordable components. Thus, the project can spread over different schools, even when the budget is restricted, with the possibility to be extended to further needs.

### 7.1 Limitations

The limitations of this project were not fully outlined due to the lack of practical application. The use of the tool in the school environment requires ethical permissions, which have yet granted.

Given the reality of some public schools in the Brazilian territory, the acquisition of the technology may still be an impediment, despite the low cost of mat components. That is because the project requires the use of a computer for the Quiz to be displayed, together with a projector or television, which are more expensive items.

### 7.2 Future work

In terms of future work, we first plan to validate our approach in a few real classes, considering students' interests, the possible impacts on the learning process, and physical measurement during the activity. New perspectives are sought with a focus on specific disciplines, through the development of new formats for the interactive floor. In the current format, the device allows for various configurations; however, it is limited to four answer elements. The software tools can also be improved, by computing a rank of students and allowing a more diverse configuration of questions, such as separating them into different disciplines. An even more audacious extension would be to map the main difficulties faced by the students while playing with the interactive floor and propose strategies to improve their learning in the discipline.

## ACKNOWLEDGEMENTS

This work was developed as a bachelor's final project and can potentially be extended by other students who may address future work and further matters related to this project. The work was originally developed in Portuguese and is still being translated into English.